\documentclass[twocolumn,showpacs,preprintnumbers,amsmath,amssymb]{revtex4}

\usepackage{graphicx}% Include figure files
\usepackage{dcolumn}% Align table columns on decimal point
\usepackage{bm}% bold math
\usepackage{textcomp} %dunno, found on internet
\usepackage{color}

\begin{document}
\newcommand{\comment}[1]{}
\newcommand{\bprime}{B^{\prime}_\rho}
\newcommand{\bdp}{B^{\prime\prime}_z}

%\preprint{XXXXXXXXX}

\title{Bose-Einstein condensation in a mm-scale Ioffe-Pritchard trap}% Force line breaks with \\

\author{Kevin L. Moore}%
\email{klmoore@socrates.berkeley.edu}%
\author{Thomas P. Purdy}%
\author{Kater W. Murch}%
\author{Kenneth R. Brown}%
\author{Keshav Dani}%
\author{Subhadeep Gupta}%
\author{Dan M. Stamper-Kurn}%
\affiliation{Department of Physics, University of California, 366 LeConte Hall \#7300, Berkeley, CA  94720}%

\date{\today}

\begin{abstract}

We have constructed a mm-scale Ioffe-Pritchard trap capable of
providing axial field curvature of 7800 G/cm$^2$ with only 10.5
Amperes of driving current.  Our novel fabrication method involving
electromagnetic coils formed of hard anodized aluminum strips is
compatible with ultra-high vacuum conditions, as demonstrated by our
using the trap to produce Bose-Einstein condensates of $10^6$
$^{87}$Rb atoms.  The strong axial curvature gives access to a
number of experimentally interesting configurations such as tightly
confining prolate, nearly isotropic, and oblate spheroidal traps, as
well as traps with variable tilt angles with respect to the nominal
axial direction.
\end{abstract}


\maketitle

%\section{\label{sec:level1}Introduction}

Magnetic traps have become a staple  of ultracold atomic physics. As
such, innovations in magnetic trapping techniques have consistently
led to new experimental breakthroughs.  For example, the invention
of the time-orbiting-potential (TOP) trap to stem Majorana losses in
spherical quadrupole traps led to the first gaseous Bose-Einstein
condensates (BECs) \cite{ande95}.  The cloverleaf trap
\cite{mewe96bec}, the QUIC trap \cite{essl98}, and other electro-
and permanent magnet configurations allowed for stable confinement
of large BECs with DC fields and variable aspect ratios; these
capabilities led, for example, to precise tests of mean-field
theories \cite{stam98coll}, observations of quasi-condensates in
reduced dimensions \cite{dett01phase}, and studies of long-lived
hyperfine coherences in two-component gases \cite{lewa02}. The
rapidly-developing magnetic-trapping technology of atom chips now
provides new capabilities for manipulating ultracold atoms and
studying their properties (e.g.\ coherence of condensates in a
waveguide \cite{wang04inter}, the decay of doubly-charged vortices
in a BEC \cite{shin04}, etc.).

A typical configuation for magnetic trapping with DC magnetic fields
is the Ioffe-Pritchard (IP) trap \cite{varenna99}.  Near the trap
center
--- at distances small compared to the size of or distance to the
magnets used to generate the trapping fields --- an IP trap is
characterized by three quantities: the axial bias magnetic field
$B_0$,  the radial quadrupole field gradient $\bprime$, and the
axial field curvature $\bdp$.  The magnitudes of these parameters
scale as $I/d$, $I/d^2$ and $I/d^3$, respectively, where $I$ is the
total current carried in the wire(s) (or magnetization of
ferromagnets), and $d$ is their characteristic length scale or
distance from the location of the magnetic trap center.  Both
because of this scaling, and because the effective radial curvature
can be greatly increased by lowering the bias field $B_0$, the
limitation to the confinement strength of an IP trap comes typically
from the maximum axial curvature which can be attained.

As indicated by the $I/d^3$ scaling of the axial curvature,
strategies for increasing the confinement of an IP trap involve both
increasing the current in the coils and decreasing the
characteristic size scale of the trap.  Magnetic traps used in most
ultracold atom experiments have been constructed on one of two
different length scales.  Centimeter (inch) scale traps, which
provide superior optical access, utilize currents of 1000's of
Amperes, typically distributed as smaller currents in each of
several turns of wire. The highest currents sustainable in such
traps, limited by resistive heating, restrict axial field curvatures
to the neighborhood of 100 G/cm$^2$.

Alternatively, magnetic confinement can be provided with modest
currents by reducing the field-producing wires and their distance to
the ultracold atoms to microscopic sizes.   This strategy has been
carried out effectively with surface microtraps
\cite{folm00,hans01,ott01chip}, resulting in versatile ultracold
atomic experiments. The typical size scale for these microfabricated
magnetic traps is $\sim$100 $\mu$m, and typically only 1 A of
current is required to produce IP traps with field curvatures in
excess of $10^4$ G/cm$^2$ \cite{hans01,curvfoot}. Microtraps are not
ideally suited for all experimental endeavors, however, as the
atomic cloud is trapped $\sim$100 $\mu$m or less from the planar
surface.

In this article we describe the design, construction, operation, and
performance of a millimeter-scale, $\sim$10 A (or $\sim$ 100
Ampere-turns) magnetic trap which bridges the two aforementioned
regimes.  This ``millitrap'' utilizes a novel fabrication scheme
which allows for the production of axial field curvatures of over
7800 G/cm$^2$ and is shown to be compatible with experimental
requirements for the creation of large BECs. We demonstrate that
this trap, owing to its high axial field curvature, allows for a
wide range of trapping geometries, ranging from the typical prolate
spheroidal to the more unusual oblate spheroidal configuration.
Further, we describe a modification of the IP trapping fields which
allows for traps with a variable tilt angle with respect to the
nominal axial direction, a capability which is compatible with
excitation of the ``scissors mode'' \cite{mara00}, the creation of
vortices \cite{madi99vort,abos01lattice} or other studies of
superfluid flow \cite{hech02irrot,stri01SFgyro} in a BEC. The trap
is also suitable for loading and trapping an ultracold atomic gas
inside a high-finesse cavity formed by conventional mm-scale mirrors
\cite{hood98,ye99trap,pinske00}(or near other mm-scale objects).

\comment{ has been used to make a 3 million $^{87}$Rb atom BEC.
The millitrap accesses a number of interesting experimental
regimes:

\begin{itemize}
\item{BECs in a tight $\{ \bar{\omega}_{\rho}$ = $2\pi\times$150 Hz, $\omega_z$ = $2\pi\times$58 Hz$\}$ trap}
\item{BECs in a nearly isotropic $\bar{\omega} = 2 \pi \times 58$ Hz trap,
suitable for precise measurements of collective excitations and the
role of quantum depletion \cite{pita98}}
\item{BECs in an oblate spheroidal trap $\{\omega_x = 2\pi\times$7 Hz,
$\omega_x = 2\pi\times$24 Hz, $\omega_x = 2\pi\times$56 Hz$\}$ trap}
\end{itemize}
}
%The curvature and anti-bias coils are
%multiple turns of thin anodized aluminum with mere $\frac{1}{2}$
%mm$^2$ cross sections. Appropriate cryogenic cooling and
%heat-sinking allows us to achieve current densities in all coils
%of nearly 200 A/mm$^2$ which, when placed properly, create the
%desired IP trap with 10$^4$ G/cm$^2$ of axial curvature. We have
%loaded 100 million $^{87}$Rb atoms at 10 $\mu$K into this trap,
%and produced a nearly pure BEC of 3 million atoms.

%\section{Experiment}

The winding pattern of the millitrap is similar to that of
inch-scale IP traps (see Figs. 1 and 2). The axial field is shaped by two pairs of
coaxial coils, with parallel currents in each pair of coils but
opposite currents in each of the two pairs. The small diameter coils
(``curvature coils'') are positioned to generate the maximum
possible curvature given their diameter.  The larger diameter coils
(``anti-bias coils'') allow for near cancelation of the large bias
field produced by the curvature coils at the trap center, while
their small axial separation allows for a slight increase (about 15
\%) in the total axial curvature. Finally, two elongated rectangular
coils (``gradient coils''), run antiparallel currents to produce a
radial quadrupole field. The dimensions of various coils were chosen
to maximize axial curvature while allowing for a 3 mm diameter
cylindrical clearance along the trap axis (for the later
accommodation of mirrors for a Fabry--Perot cavity), and a 1 mm
clearance along the radial directions for the purpose of imaging.
Further details on the positioning and cross sectional area of the
coils are shown in Table \ref{tab:params}.

\begin{figure}
\includegraphics[angle = 0, width = 0.5\textwidth] {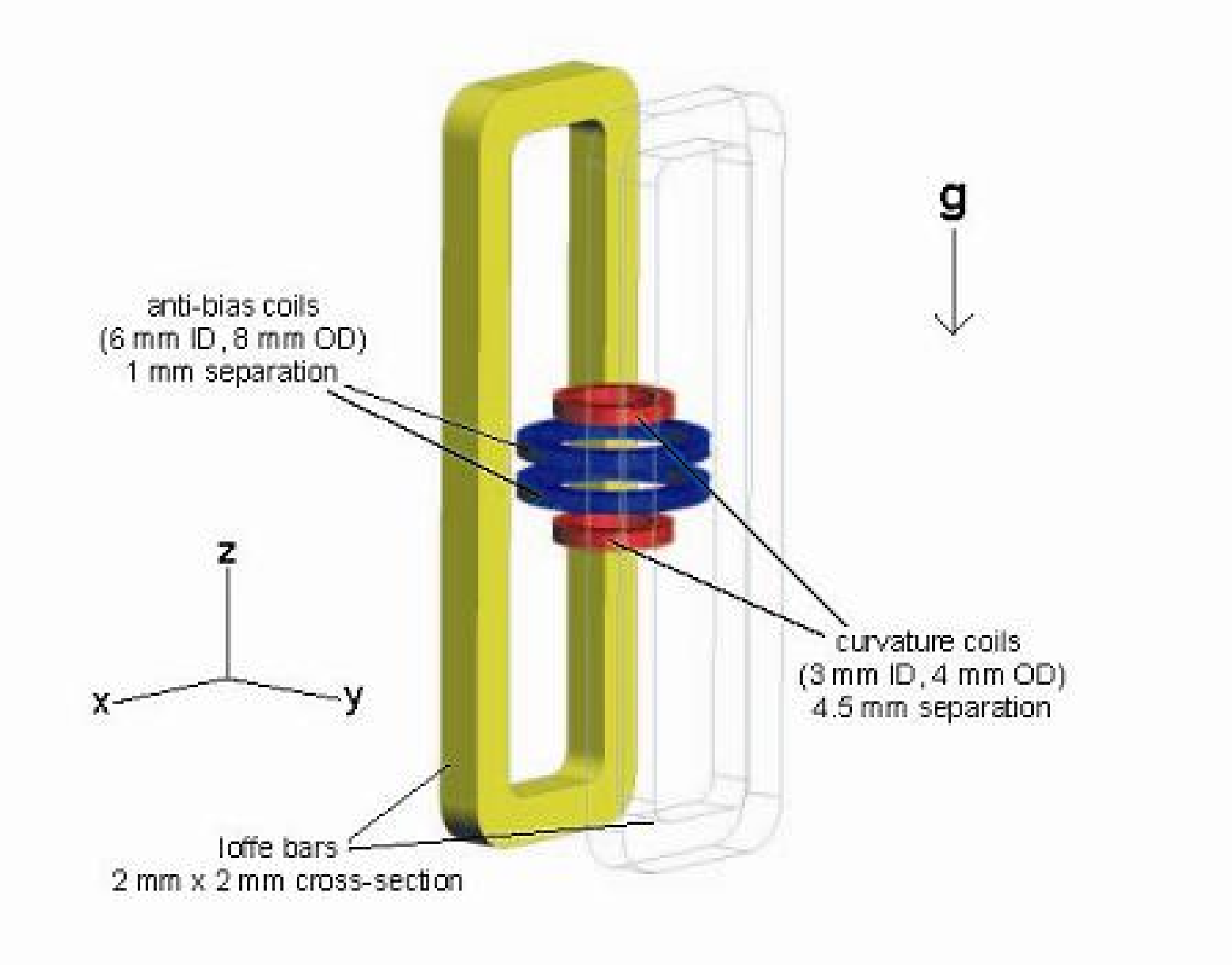}
\caption{\label{fieldsdiagrams} Sketch of the mm-scale IP trap. The
primary curvature coils (red), the anti-bias coils (blue), and the
gradient coils (yellow) are depicted in this diagram as solid
bodies, but are in actuality multiple turns of wire with protruding
leads. For clarity the coil leads have been omitted and the nearest
gradient coil is shown as transparent.  \emph{Higher resolution
version of figures at
http://physics.berkeley.edu/research/ultracold.}}
\label{fig:3dsketch}
\end{figure}

%\subsection{Trap design}

%When the design for this trap began, the only geometrical
%constraint on the design was the requirement that the radius of
%the curvature coils to be no less than 1.5 mm
%\cite{size:footnote}. Once the coil radius \emph{a} is determined,
%the optimal distance \emph{$z_o$} from the coil to the trap center
%is located at $z_o = \frac{\sqrt{3}}{2} a$, given by optimizing
%the expression for the field curvature:

%\begin{equation}
%\frac{d^2 B_z}{dz^2} = 6 \mu_o I
% \frac{a^2(a^2-4z_o^2)}{(a^2+z_o^2)^{\frac{7}{2}}}
%\end{equation}

%The anti-bias coils are placed so as to add to the field curvature
%as well as serve their primary function of reducing the bias field
%produced by the curvature coils (see Fig. 1b).  Given the minimum
%required cross section for these coils as well as the desire for
%optical access to the atomic cloud from the side, the anti-bias
%coils are constructed with an outer diameter of 4 mm.  As they are
%positioned in this trap the anti-bias coils will add 15$\%$ more
%confining field curvature.  Finally, the Ioffe bars are placed as
%close as possible in the plane perpendicular to the symmetry axis.

%\subsection{Trap construction}

To maximize the current density while avoiding large input
currents and uncontrolled magnetic fields from current leads,
multi-turn coils (with total cross sections on the order of 1
mm$^2$) were used.  The maximum current density attainable in
coils fabricated by various methods is limited by the steady state
temperature of the coils, due to the tendency of the coil
resistance to rise with temperature. We found that, for all
implementations, there is a threshold at which no more current can
be added to a coil without the resistance increasing exponentially
from overheating.  Thus, in order to minimize resistive heating
and maximize heat dissipation, it is desirous to choose a
fabrication method which allows for the cross-sectional area to be
efficiently packed with current carrying conductor rather than
electrical (and typically thermal) insulation.

Guided by these criteria, we chose to form electromagnetic coils
from multiple turns of hard-anodized pure aluminum foil strips. The
assembly procedure is illustrated in Fig. 2.
Shear-cut strips of aluminum foil were cleaned and then hard
anodized in sulfuric acid after smoothing their jagged edges with
lubricated fine grit sandpaper.  The thickness of the insulating
Al$_2$O$_3$ layer (on the order of microns) was controlled by
varying the duration of the anodization, and chosen to be thick
enough to reliably prevent current shorts between turns of the coil
but thin enough to allow the coils to be wound without fracture.
Coils were then wound on Teflon mandrels with a UHV-compatible,
thermally-conductive epoxy applied between turns. The epoxy was set
by baking the coil and mandrel at 150$^\circ$C for two hours, after
which the coil was removed  and then tested for electrical shorts
through both DC resistance measurements and AC magnetic field
measurements.

The coils were then inserted into a compound mounting and
heat-sinking structure and secured by epoxy (curvature and antibias
coils) or by pressure (gradient coils). Portions of the mount in
contact with the coils were formed from hard-anodized aluminum.
Current connections to the coil were formed by removing oxide layers
from the leads and then clamping them tightly between two pieces of
copper.  Finally, the trap and mounting structure was installed in a
UHV vacuum chamber, with current connection made through
polyimide-insulated copper wires to a set of 20 A vacuum current
feedthroughs.  The mounting structure also contains two hollow
channels for circulation of liquid nitrogen.  Operating the magnetic
trap at liquid nitrogen temperatures lowers the resistance of the
aluminum coils by a factor of four compared to that at room
temperature, allowing higher current densities to be maintained.
 Following a bakeout of the millitrap at a temperature of
250$^\circ$C, lifetimes of over 100 s were observed for atoms
trapped in the millitrap, demonstrating the vacuum compatibility of
all materials used in its construction.

\begin{figure*}
\includegraphics[angle = 0, width = \textwidth]{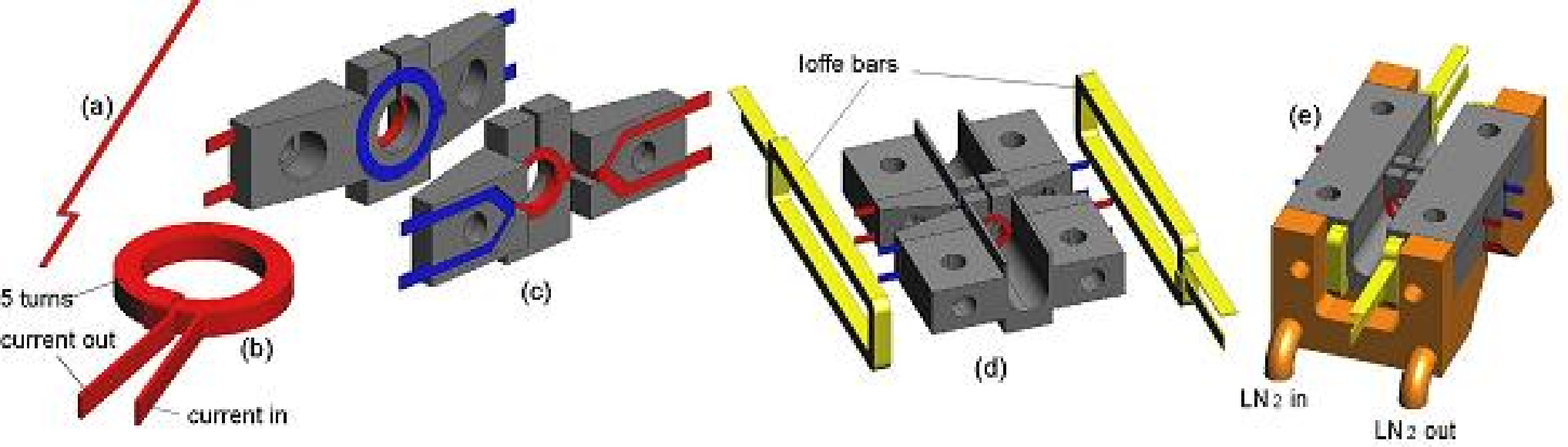}
\caption{\label{fig:assembly} Sketch of the assembly procedure and
part integration.  (a) Aluminum strips were cut with a z-shaped
pattern to allow for the extraction of the interior current channel
after the coil is wound.  (b) A curvature coil with input and output
current leads.  (c) Curvature coils (red) and anti-bias coils (blue)
were epoxied into anodized aluminum (grey) faceplates; current leads
protrude from sides.  (d) The faceplates were attached to an
anodized aluminum mount which allows the gradient coils (yellow) to
slide over the assembly.  (e) A top fixture plate holds the mount in
place by bolting into a copper mount (orange) below.  A closed path
inside the copper piece allows liquid nitrogen to be circulated.}
\end{figure*}

\begin{table*}
\caption{\label{tab:params}Parameters for aluminum coil windings.}
\begin{ruledtabular}
\begin{tabular}{cccccccc}
Coil & Inner Diam. & Outer Diam. & Foil thickness & Width & Cross-section & No. turns & Heat generated @ 10 A\\
\hline curvature & 3 mm & 4 mm & 0.006 in. & 1 mm & 0.5 mm$^2$ &  5 & 2 Watts\\
anti-bias & 6 mm & 8 mm &  0.008 in. & 0.75 mm & 0.75 mm$^2$ & 4 & 2 Watts\\
gradient & N/A & N/A &  0.008 in. & 2 mm & 4 mm$^2$ & 9 & 10 Watts\\
\end{tabular}
\end{ruledtabular}
\end{table*}

%\subsection{Trap Testing}

To provide the most flexibility in operating the millitrap, separate
electrically-floating power supplies were used for each coil.  Also
included in the electrical setup were a set of inductor-capacitor
filters and an interlock system to protect the millitrap from
overheating. Electrical characterization of the millitrap following
the vacuum bakeout revealed several undesired low-resistance
(several Ohm) connections between different coils, indicating
electrical connections through the common mounting structure.  These
inter-coil connections should have no effect since independent
supplies are used for each coil. The possible presence of undesired
intra-coil connections, e.g.\ connections between turns on the
multiple-turn coils, was tested by measuring parameters of magnetic
traps formed with varying currents in each of the curvature,
anti-bias, and gradient coils. No clear evidence for such flaws was
obtained.

%\section{Performance}

%After the vacuum chamber has been baked out with the trap in place,
%cold atoms must be delivered to the IP trapping region.  Because of
%the small length scales obscuring the optical access to the trapping
%region (1 mm in the axial direction) it is unreasonable to imagine a
%large magneto-optical trap (MOT) population being collected in the
%trapping region itself. Similar to some surface microtrap
%experiments, large MOT coils external to the vacuum system are
%employed to obtain an initially large collection of atoms that is
%subsequently transferred into the main magnetic trap.

\begin{figure*}
\includegraphics[angle = 0, width = 0.9\textwidth] {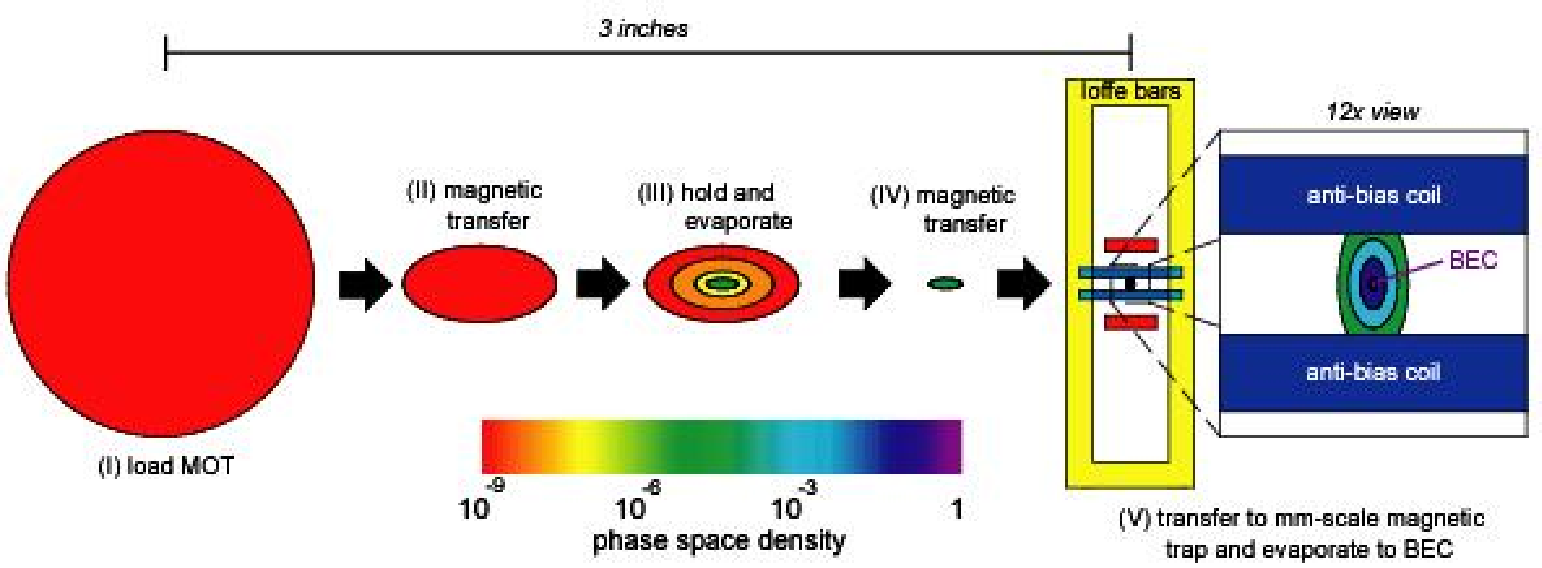}
\caption{\label{exptsketch} Sketch of experimental sequence.  (I)
Atoms are loaded into the MOT and subsequently trapped in a
spherical quadrupole trap.  (II) The atoms are then transferred 1.75
inches towards the mm-scale IP trap and (III) evaporated to a phase
space density of $\Gamma \sim 10^{-5}$.  (IV) The cloud is
magnetically transferred into the IP trap and (V) captured by a
curvature coil and an anti-bias coil in a spherical quadrupole trap.
The full millitrap is then turned on and the atoms are confined in
the IP field with a 2 G bias field.  The cloud is then further
evaporated, forming a pure BEC of 1 million atoms.}
\label{fig:setup}
\end{figure*}

Cold atoms were loaded into the millitrap by optically cooling and
trapping atoms in one portion of the UHV chamber, and then
magnetically transporting them to the millitrap region.  This
multi-stage experimental procedure is depicted in Fig. 3.  In a ``loading region'' which is displaced 3
inches horizontally from the millitrap, a $5\times 10^9$ atom MOT
was loaded from a Zeeman slowed beam of $^{87}$Rb. About $2 \times
10^9$ atoms were trapped in the $|F=1, m_F=-1\rangle$ magnetic
sublevel by a spherical quadrupole magnetic trap with an axial
gradient of 200 G/cm. The atoms were then transported using two sets
of stationary anti-helmholtz coil pairs external to the chamber
(similar to Ref. \cite{grei01trans}), one of which is centered at
the ``loading region'' and which is used for the initial spherical
quadrupole trap and the other centered at the millitrap. As the two
anti-helmholtz coil pairs overlap each other, the
magnetically-trapped cloud was easily transported between the two
coil centers by varying the currents in the two quadrupole coil
pairs. During this transport, the atomic cloud was cooled by RF
evaporation to reduce the cloud size to about 400 $\mu$m before
passing the atoms through the 1 mm gap between the millitrap
anti-bias coils.

%The magnetically-trapped atoms are transported 1.75 inches toward
%the millitrap. At this position both external quadrupole coils are
%running full current, forming a quasi-quadrupolar field $\vec{B}$
%= $\{$(100 G/cm)x, (150 G/cm)y, -(250 G/cm)z$\}$. The atoms are
%then evaporatively cooled to 15 $\mu$K, reducing the size of the
%cloud to 400 $\mu$m, before transporting them to the center of the
%millitrap. This is crucial because the anti-bias coils in the IP
%trap allow only a 1 mm gap through which the atoms are threaded.
%The cloud is cooled to 15 $\mu$K, at which point the cloud is only
%400 $\mu$m across yet not so cold that the Majorana loss rate is
%sizeable. The atoms are then transported the remaining 1.25 inches
%into the IP trap center.

%The IP trap is oriented with the axial field collinear with
%gravity. The axial field curvatures are large enough that the
%sagging of the cloud due to gravity is very small (35 $\mu$m at
%10$^4$ G/cm$^2$).

Transfer of the atoms from the external-coil-based spherical
quadrupole trap to the IP trap was accomplished in two stages of
``handshaking.''  First, atoms were transferred to a spherical
quadrupole trap formed by two of the six millitrap coils (a
curvature coil and an opposing anti-bias coil); at 2 A running
through each of these coils, a quadrupole trap with 150 G/cm axial
gradient was produced, nearly matching the field strength generated
by 400 A of current running through the external quadrupole coils.
The spherical quadrupole trap was then suddenly (within $100$
$\mu$s) replaced with the IP millitrap. This sudden quadrupole-to-IP
transfer caused 25$\%$ (or less) of the atoms to be lost. RF
evaporative cooling was then performed in a prolate IP trap, with
trapping frequencies of $(\omega_x,\omega_y,\omega_z) = 2 \pi \times
(151,138,52)$ Hz (axes oriented as in Fig. 1),
yielding atomic clouds near or below the BEC transition temperature
(about 300 nK for our system). The transition temperature was
reached with $2.5 \times 10^6$ atoms, and nearly pure condensates of
$1 \times 10^6$ atoms produced upon further cooling.

The strongest confinement provided by the millitrap depends on
whether such confinement is provided for long or for short trapping
times.  For example, up to about 7 A of current can be maintained in
the curvature and anti-bias coils on a steady-state basis. Coils
were safely operated at higher currents, up to about 11 A, although
we found that after about 100 ms, the resistive heating of the coils
led to increased outgassing which worsened the vacuum conditions in
the millitrap region.  The axial curvature provided under these
conditions was measured in-situ using the trapped atoms as a probe,
both by measuring the axial oscillation frequency of the trapped
cloud, as well as by measuring the axial displacement of the cloud
due to the application of a known axial field gradient.  From these
measurements, we determine that steady-state axial curvatures of
5300 G/cm$^2$ (7 A setting) and brief confinement with 7800 G/cm$^2$
(10.5 A setting) can be reached.   Gradient coils are operated at a
maximum of 11 A, yielding radial gradients of 220 G/cm.

One unexpected feature of this strong IP trap is a remarkably high
efficiency of RF evaporation.  This efficiency can be quantified by
comparing the factor gained in phase space density $\Gamma$ through
the evaporative cooling loss of a given factor in atom number $N$,
obtaining, e.g.\ a figure of merit $f = - d\ln\Gamma / d\ln N$, with
$\Gamma$ and $N$ parameterized along some evaporation trajectory.
Typical figures of merit cited in the literature for evaporation
from IP traps are $f=2$ to $f=3$ \cite{arlt99,kett96evap}.  In our
mm-scale IP trap, a factor of over $10^5$ in phase space density is
efficiently gained by evaporative cooling to the Bose-Einstein
condensation transition temperature with an overall figure of merit
of $f=4.5$.

To account for this high efficiency, we note that the IP trap, aside
from being strongly confining and thus compressing atomic clouds to
high collision rates, is also nearly isotropic. We suspect that the
condition of near isotropy improves the efficiency of evaporative
cooling relative to that in the typically-used anisotropic traps
since high-energy atoms produced collisionally in the gas can easily
escape the center of the cloud in \emph{any direction}, and thereby
reach the trap boundary established by the applied RF radiation.  In
contrast, in a cigar-shaped cloud with high aspect ratio, the large
axial collisional depth can prevent the escape of all high-energy
atoms except those travelling nearly purely in the radial direction.
Further, we note that high evaporation efficiency is obtained in our
trap in spite of the vertical orientation of the axial direction; in
contrast, IP traps with weaker axial confinement are rarely oriented
in this manner so as to avoid the onset of lower dimensional
evaporation due to gravitational sag \cite{kett96evap, pink98}.

We have investigated several new features which are afforded by the
large axial curvature in our trap.  For instance, considering the
generic magnetic field configuration of an IP trap and expanding
about the minimum of the magnetic field, an effective radial
curvature is obtained as $B^{\prime\prime}_\rho = {\bprime}^2 / B_0
- \bdp/2$. The dependence of the radial trap strength on the applied
bias field $B_0$ offers a simple means of varying the aspect ratio
of the trap arbitrarily, ranging from prolate ($\bdp > B^{\prime
\prime}_\rho$) to near-isotropic ($\bdp \simeq B^{\prime
\prime}_\rho$) to oblate ($\bdp < B^{\prime \prime}_\rho$)
geometries. While experiments using IP traps have typically employed
prolate or near-isotropic geometries, the oblate geometry has been
avoided since the very weak confinement afforded by such traps
(limited to below the already weak axial confinement), makes it
difficult to compensate for gravitational sag and stray magnetic
fields.  Thus, by greatly boosting the typical axial confinement
strength, our trap gives more convenient access to oblate DC
magnetic traps, with advantages for the study of two-dimensional
\cite{gorl01lowd,rych04,smit04twod} and/or rotating condensates.

Fig. 4 shows the range of trapping geometries
accessed by our millitrap.   After evaporatively cooling a thermal
gas to a temperature of about 500 nK, the bias field $B_0$ was
ramped to values ranging from $2$ G to $18$ G while holding the
axial curvature at $\bdp \simeq 4000$ G/cm$^2$ and radial gradient
at $\bprime = 205$ G/cm. We then displaced the cloud slightly in
this new trap configuration, and recorded the harmonic motion of the
trapped cloud to determine trap frequencies along three orthogonal
directions. For this purpose, absorption imaging was employed along
either of two imaging axes --- one through the 3 mm vertical
aperture along the vertical trap axis, and the other along the
horizontal $\hat{y}$ direction through the 1 mm gap between the
anti-bias coils.

These measurements illustrate the breaking of radial trap symmetry
in our trap due to gravity.  This can be understood by considering
that the atomic cloud sags under gravity to the point where the
axial gradient of about 30 G/cm gives a force on atoms in the
$|F=1,m_F=-1 \rangle$ equal to the gravitational force. By the
condition $\vec{\nabla} \cdot \vec{B} = 0$, the presence of this
axial gradient implies a radial field gradient of 15 G/cm which
breaks the symmetry of the radial quadrupole field, adding to the
magnetic field gradient along one direction ($\hat{y}$) while
subtracting from that along the other direction ($\hat{x}$). Thus,
triaxial, rather than cylindrically symmetric, traps are produced.

One motivating factor in our tailoring the aspect ratio of the IP
trap is the desire to detect the presence of quantum depletion by
precise measurements of collective excitation frequencies, as
proposed by Stringari and Pitaevskii \cite{pita98}. If one considers
a fixed condensate number and axial trap strength, one finds that
the largest magnitude frequency shift of the lowest collective mode
would be obtained with traps that are nearly isotropic; even though
higher condensate densities (and hence higher quantum depletion) are
produced in prolate traps, the quadrupole modes in this case are
more surface-like, rather than compressional, in character, and
hence are only weakly affected by depletion effects.   In our case,
the broken symmetry due to effects of gravity produced, at best,
nearly isotropic traps. For instance, Fig. 5
shows time-of-flight absorption images of atoms from a
$\omega_x:\omega_y:\omega_z = 0.91 : 1.08 : 1.00$ trap. The familiar
pronounced anisotropy of an expanding BEC is absent from such images
due to the trap isotropy.

\begin{figure}
\includegraphics[angle = 0, width = 0.5\textwidth] {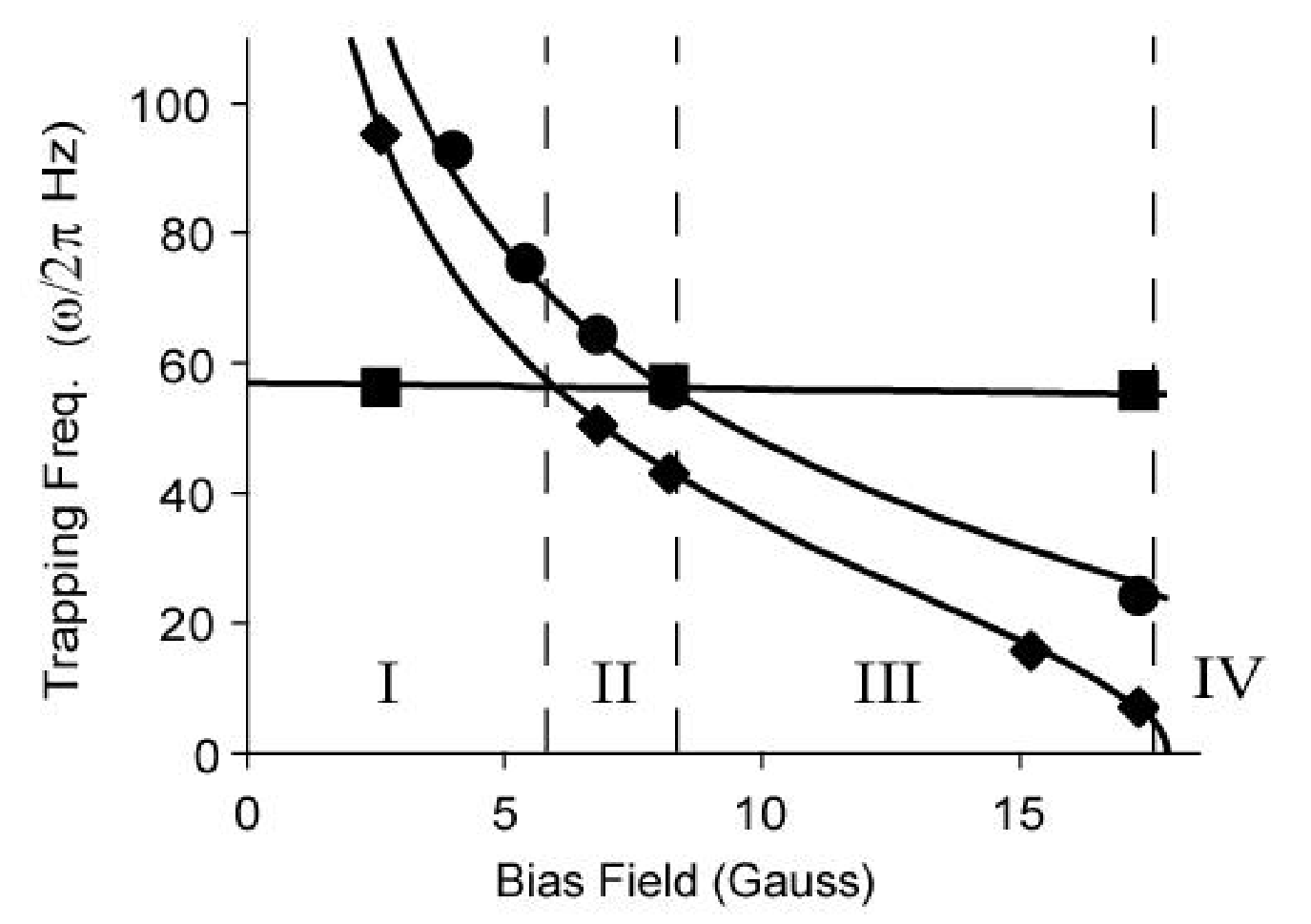}
\caption{\label{Trap Frequencies} Measured axial trapping frequency
$\omega_z$ (squares) and transverse trapping frequencies $\omega_x$
(diamonds) and $\omega_y$ (circles) as a function of bias field,
which was controlled by varying the current in the anti-bias coil
pair.  The solid lines are theoretical predictions for the trapping
frequencies. The only free parameter in the transverse trapping
frequency fit is the gradient coil contribution which was allowed to
vary within its measured uncertainty. Four distinct regimes can be
identified: I - the prolate spheroidal regime (``cigar''-shaped
clouds), II - the nearly-isotropic regime, III - the oblate
spheroidal regime (``pancake''-shaped), and IV - the unstable
regime.}\label{fig:trapfreq}
\end{figure}

\begin{figure}
\includegraphics[angle = 0, width = 0.5\textwidth] {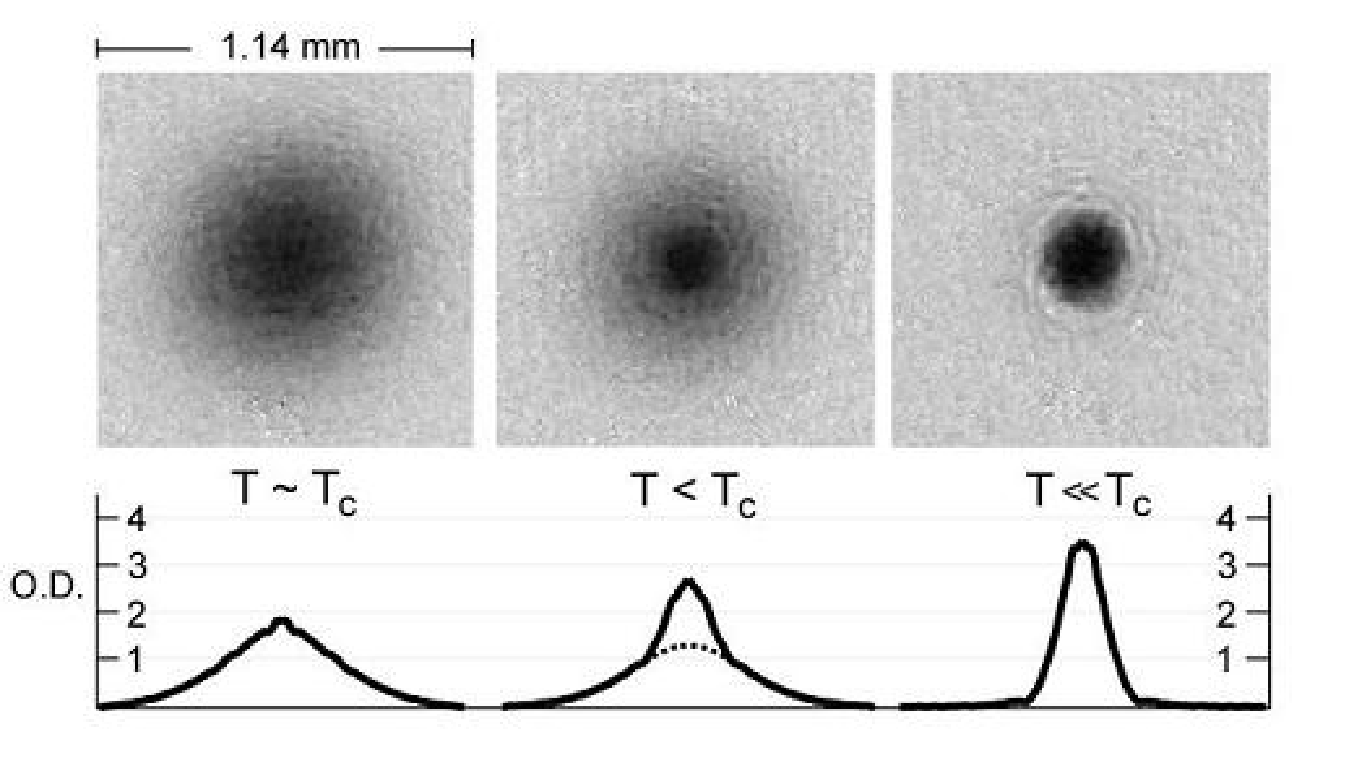}
\caption{\label{BEC pics} Absorption images of a nearly-isotropic
ultracold gas. Images show separate 36 ms time-of-flight images of a
thermal cloud ($1.5 \times 10^6$ atoms), bimodal distribution, and
pure BEC ($0.5 \times 10^6$ atoms), respectively. The trapping
frequencies for this trap are $\{ \omega_x,\omega_y, \omega_z \} =
2\pi\times \{ 52, 62, 57 \}$Hz.  Below the images are associated
radial averages of the optical densities.  The bimodal distribution
(center plot) is clearly seen with the condensate rising from the
Gaussian fit to the thermal wings (dotted line).}
\label{fig:becexpand}
\end{figure}

\comment{
\begin{table}
\caption{\label{tab:table2}Phase space trajectory for producing
Bose-Einstein condensates in the millitrap.  Starting from a
Zeeman-slower-loaded MOT, atoms are captured in a spherical
quadrupole magnetic trap provided by external coils, evaporatively
cooled, transferred into the IP millitrap, and then cooled further
to quantum degeneracy.  Detailed trap parameters are provided in the
text.}

\begin{ruledtabular}
\begin{tabular}{cccc}
Trapping Field & Atom Num. & Temp. & Phase Space Density\\
\hline
MOT & 5$\times10^9$ & 100 $\mu$K & --- \\
Ext. quad coils\footnotemark[1] & 2$\times10^9$ & 400 $\mu$K & 10$^{-9}$\\
Ext. quad coils\footnotemark[2] & 55$\times10^6$ & 15 $\mu$K & 2$\times$10$^{-5}$\\
Mini quad coils & 55$\times10^6$ & 12 $\mu$K & 2$\times$10$^{-5}$\\
Mini IP trap\footnotemark[1] & 50$\times10^6$ & 20 $\mu$K & 10$^{-5}$\\
Mini IP trap\footnotemark[2] & 6$\times10^6$ & 400 nK & 1\\
\end{tabular}
\end{ruledtabular}
\footnotetext[1]{Before evaporative cooling.} \footnotetext[2]{After
evaporative cooling.}
\end{table}
}

Another feature highlighted by the large axial confinement of our
trap is a means to vary the orientation of the trap with respect
to the nominal axial direction.   This effect arises from
considering the effects of displacing the radial quadrupole
gradient field so that its zero-field axis no longer coincides
with the axis of the curvature fields.  That is, one considers the
fields
\begin{eqnarray}
\vec{B}_{curv} & = & B_0 \hat{z} + \frac{ \bdp}{2} \left[ \left(z^2 - \frac{x^2 + y^2}{2}\right) \hat{z} - z \left(x \hat{x} + y \hat{y} \right)\right] \nonumber \\
\vec{B}_{grad} & = & \bprime \left[ (x - x_0) \hat{x} - (y - y_0)
\hat{y} \right] \label{eq:displacedip}
\end{eqnarray}
where $(x_0, y_0)$ is the position of the gradient-preferred-axis in
the $\hat{x}$ -- $\hat{y}$ plane.  This position is controlled
experimentally by applying uniform radial fields to a well-aligned
($x_0 = y_0 = 0$) IP trap.  Such misalignment yields both a variable
displacement and variable tilt of the resulting magnetic trap, which
can be understood as follows.  Considering for now just the $z = 0$
plane, the location of the magnetic trap is determined by the
competition between $\vec{B}_{grad}$, which tends to locate the
cloud at $(x_0, y_0)$, and the radial variation of the axial field
$\vec{B}_{curv}$ which, for small displacements, exerts a radially
repulsive force.  The $(x,y)$ position of the resulting field
minimum varies for $z \neq 0$ due to the fact that the
$\vec{B}_{curv}$ fields now acquire radial components, displacing
the position of the radial-field minimum from $(x_0, y_0)$.

To illustrate this effect, we present in Fig. 6
the tilt angles $\theta$ with respect to the $\hat{z}$ axis of the
weakest trap axis in a prolate IP trap, as derived from Eqs.\
\ref{eq:displacedip}. Field parameters of $\bdp = 2000$ G/cm$^2$,
$B^{\prime}_{\rho} = 180$ G/cm, and displacements $y_0 = 0$ and
variable $x_0$ are chosen to match experimental settings. The tilt
angle varies over a wide range of $x_0$, out to a limiting
displacement $x_c = 2 \sqrt{B_0/\bdp}$ beyond which a non-zero-bias
trap is no longer produced.

\begin{figure}
\includegraphics[angle = 0, width = 0.5\textwidth] {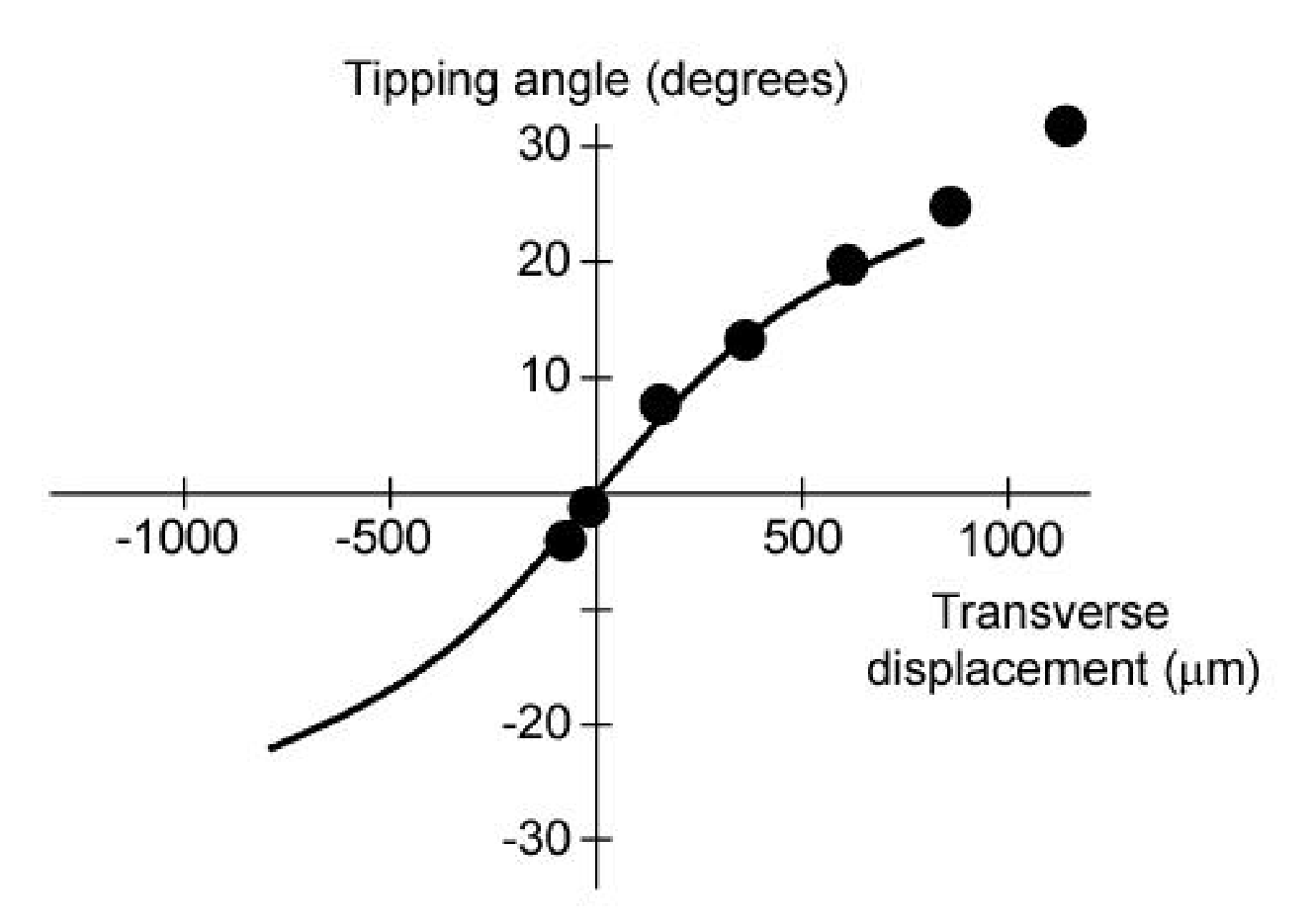}
\caption{\label{fig:anglefigure}Tilting atomic clouds in an IP
magnetic trap.  \emph{In situ} images of tipped prolate clouds yield
both the displacement (distinct from $x_o$) and the tipping angle (data shown as points).
These data are compared with calculations (solid line) obtained
from the generic IP field expressions of Eqs.\ \ref{eq:displacedip}
for the trap parameters of this experiment ($\bdp =
2000$ G/cm$^2$, $\bprime = 180$ G/cm, and $y_0 = 0$).  The theoretical curve is shown only over the range of displacements at which the IP traps (non-zero bias fields) are retained.  Beyond this range, the displaced traps become filled, asymmetric spherical quadrupole traps, as presumably applies to the two data at highest displacements.}
\end{figure}

\comment{, displacement $|x_0|
> x_c = 810 \, \mu$m would no longer support a non-zero bias trap}

Aside from varying the tilt angle, this variation of the IP trap
also changes the trap frequencies.  Indeed, we observed
experimentally that the ``axial'' trap frequency, i.e.\ the smallest
frequency in a prolate IP trap, can be dramatically reduced in the
case of a misalignment
($x_0$ and/or $y_0 \neq 0$). This leads to an
apparent discrepancy between this trap frequency, which was
determined by following the oscillatory motion of a trapped cloud,
and a measurement of $\bdp$, as determined from measuring the upward
($\hat{z}$) displacement of the magnetic trap for a given axial
field gradient, when the misalignment was large. Once external
fields were applied to correct this misalignment, the measured trap
frequencies and axial field curvatures were in agreement.

In conclusion, we have constructed a novel mm-scale IP magnetic trap
which provides the means for tailoring magnetic potentials on length
scales intermediate to the larger, inch-scale electromagnets and
smaller microfabricated devices.  The millimeter length scale is in
some ways natural for manipulating cold atomic clouds, generating
sufficiently deep and well behaved potentials over the $\sim$100
micron scale of typical gaseous samples.  This trapping technology
may thus provide a flexible means to transport ultracold clouds or
construct large scale waveguides appropriate for condensate-based
interferometry schemes
\cite{shin04ifm,wang04inter,hans01trappedIFM,ande02}. Further,
making use of the strong axial confinement of the millitrap, we have
demonstrated a wide range of trapping geometries which may enable a
variety of experiments.  For instance, the ability to continuously
manipulate the tilt of a cigar-shaped condensate with respect to a
fixed axis, simply by the application of uniform magnetic fields,
provides a new all-magnetic method for imparting angular momentum to
a trapped gas. Compared with laser-based excitation schemes, the
utility of which is limited by the length scales of an optical focus
(Rayleigh range, beam waist radius) \cite{madi99vort}, this method
may allow the excitation of vortices in BECs with extremely small
radial dimensions. Finally, the achievement of large BECs in the
millitrap, which by design is compatible with existing technologies
for high-finesse Fabry-Perot optical resonators, accomplishes a
significant milestone toward the application of cavity quantum
electrodynamics to magnetically trapped ultracold atoms.

%\section{Results}

%\section{Discussion}

%\begin{acknowledgments}

We acknowledge the skillful work of Dave Murai and Armando Baeza of
the UCB Physics Machine Shop in constructing and installing the
mount pieces. The authors' effort was sponsored by the Defense
Advanced Research Projects Agency (DARPA) and Air Force Laboratory,
Air Force Materiel Command, USAF, under Contract No.
F30602-01-2-0524, the NSF (Grant No.\ 0130414), the Alfred P. Sloan
and David and Lucile Packard Foundations, and the University of
California.  KLM acknowledges support from the National Science
Foundation. SG acknowledges support from the Miller Institute for
Basic Research in Science.

%\end{acknowledgments}

\bibliographystyle{prsty}
%\bibliography{allrefs}

\begin{thebibliography}{10}
\bibitem{curvfoot}
Specialized atom-chip traps have been fabricated which may give
axial
  curvatures up to $10^8$ G/cm$^2$ \cite{lev04nano,drnd98}  .
\end{thebibliography}
%\allrefs

\end{document}